Torin K. Clark, Aaron A. Allred, Jaime Mateus, Marissa Rosenberg

# Spaceflight-associated Motion Sickness in Commercial Crew Astronauts

## Abstract

Spaceflight-associated motion sickness, including space motion sickness (SMS) and terrestrial readaptation motion sickness (TRMS), has presented as the most operationally limiting effect on the human in the days following a gravity transition. Published data suggest that astronaut head movements provoke symptoms, but those studies rarely controlled for them. Most previous studies use postflight retrospective reporting and have not quantified temporal dynamics or persistence of symptoms; no study has published motion sickness in a cohort of commercial flyers. Here, we comprehensively evaluate SMS and TRMS within commercial flyers (N=8; 4F, 4M) for the first time, across two commercial missions of short duration (Polaris Dawn and Fram2). Utilizing a standard reporting methodology, we find an 87.5% incidence of SMS, which persisted for at least three days on orbit in 25% of crewmembers. The symptoms included vomiting in 62.5% of crewmembers. Regarding TRMS, incidence was 50%, severity was significantly lower than SMS (p = 0.007), and it tended to persist for a shorter duration than SMS for these short-duration flights. These findings establish a systematic baseline for spaceflight-associated motion sickness, informing its operational impact and salience as human spaceflight missions expand in accessibility and scope.

## Introduction

Astronauts often experience symptoms of motion sickness shortly after gravity transitions, including transitions from Earth to microgravity, then again upon returning to Earth following their missions. As other physiological effects of altered gravity either occur over longer time periods or are less noticeable, symptoms of motion sickness are often the most salient bodily response experienced by crewmembers over the first days of their mission. Spaceflight-associated motion sickness can be separated into 1) symptoms experienced upon entering microgravity, termed space motion sickness (SMS) and 2) symptoms experienced upon returning to Earth after neurovestibular adaptation to microgravity (Clark, T.K., 2019; Young et al., 1984a), recently termed terrestrial readaptation motion sickness (TRMS). In prior decades, the nomenclature of “space adaptation syndrome” (SAS) was used synonymously with SMS, but more recently SAS has been used to refer to the broader set of acute physiological changes associated with weightlessness. TRMS has also been interchangeably referred to as entry motion sickness (EMS) (Ortega & Harm, 2019; Reschke et al., 2017) or “mal de barquement that occurs after re-entry” (Lackner & DiZio, 2006). Because motion sickness may occur following a transition from microgravity to any gravity-rich environment, TRMS distinguishes between that which occurs following return to Earth and motion sickness that may be anticipated following landings on the Moon or Mars.

Currently, it is thought that SMS and TRMS are not unique etiological entities from general ‘motion sickness’ as experienced on Earth (Lackner & DiZio, 2006; Oman et al., 1986), but rather represent forms of motion sickness elicited by particular provocative environments, similar to carsickness and seasickness (Golding, 2016; Oman, 1990; Yates et al., 1998). Symptoms typically include pallor, increased body warmth, cold sweating, dizziness, drowsiness, and, in more severe cases, nausea and vomiting (Oman et al., 1986; Reschke et al., 1998, 2017). These symptoms have been reviewed elsewhere (Lackner and Dizio 2006), appear similar for TRMS (Reschke et al., 2017), and largely resemble those of other forms of prolonged motion sickness (though in SMS, some symptoms may be more or less common (Oman et al., 1986), which may be due to the interaction of other, concurrent physiological changes (Lackner & DiZio, 2006)). There is also recognition of the associated "sopite syndrome" referring to chronic drowsiness, fatigue, mood and personality changes, and the lack of initiative (Lackner & DiZio, 2006).

Operationally, SMS impacts mission activities over the first handful of days on orbit. For example, due to the risk of vomiting in the spacesuit leading to asphyxiation or aspiration, extravehicular activities (EVAs, or space walks) are not scheduled by NASA until after 72 hours on-orbit (Davis, Jennings, Beck, et al., 1993; Ortega & Harm, 2019). During the Shuttle era, minimum duration flights have been defined to be at least 3 days to help permit recovery from SMS before entry and landing (Davis, Jennings, Beck, et al., 1993). Further, crewmembers are typically cross-trained so that early critical activities are not dependent upon a single crewmember who may be impaired from SMS (Jennings, 1998). For TRMS, the operational impacts depend upon the reentry mode and the astronauts' role/responsibilities during and following landing (e.g., pilot landing the Shuttle or a crewmember attempting a postlanding emergency egress (Bacal et al., 2003)). For example, it was recommended that Commanders with prior Shuttle landing experience be named to extended-duration flights (Jennings, 1998).

Cases of SMS, and to a lesser extent TRMS, have been reported and studied for decades (see (Heer & Paloski, 2006) for a concise summary, (Lackner & DiZio, 2006) for details, and (Oman, 1998) for a theoretical review). Reviewing the existing evidence regarding SMS and TRMS, we will focusing on peer-reviewed published studies (except when noted), quantifying the *incidence* (i.e., the proportion of astronauts reporting symptoms), *severity* (i.e., the intensity of those symptoms, including vomiting), and *persistence and temporal dynamics* (i.e., how long following a gravity transition did symptoms remain, as well as how quickly did they develop), including sufficient methodological details to allow for comparison to the current study. As has been noted (Lackner & DiZio, 2006; Oman et al., 1986), under spaceflight operational conditions, it is more difficult to assess motion sickness symptoms and severity due to uncontrolled crewmember tasking and activity levels, different approaches to anti-motion sickness medications (which may not be documented), environmental conditions, and vehicle motion.

For SMS, the first documented case was on Vostok-2 in 1961, humanity's first full day in space (described in (Graybiel et al., 1977)), and early Soviet crewed flights identified SMS as a common occurrence (Matsnev et al., 1983). In early NASA flights, there was not a systematic investigation of SMS, but anecdotally, there were no reported cases during the Mercury (0 of 6 astronauts reported symptoms of SMS) or Gemini (0 of 16) programs (Berry & Homick, 1973). The first clear-cut cases were during Apollo (Berry & Homick, 1973; Graybiel et al., 1977), where "data were gleaned from a variety of sources and cross-checked" (Berry & Homick, 1973) finding 9 of 25 flyers (Apollo missions 7-15, 11 who were first time flyers, ages 33-47, all male) reported at least stomach awareness (incidence of 36%), while in terms of severity, 4 reported nausea and 3 crewmembers vomited. Temporal dynamics or persistence were not captured during Apollo except noting that SMS lasted "up to five days in the case of one crewman" (Berry & Homick, 1973). The authors also noted that head movements "appear to produce motion sickness difficulty in flight".

The first prospective, systematic investigation of SMS was on Skylab (initial results published in 6 crew across 2 Skylab missions (Graybiel, 1975), then expanded to 9 astronauts across all 3 Skylab missions (Graybiel, 1980), and more details can be found in the NASA report (Graybiel et al., 1977)). The crew (ages 36-43, all male) reported symptoms using a standardized symptom questionnaire (Graybiel et al., 1968) as well as anti-motion sickness medications taken during their flights. In terms of incidence, 5 of 9 crewmembers reported SMS symptoms (56%), with 2 vomiting. The symptom reports were scored to capture severity and temporal dynamics, noting one crewmember experienced motion sickness "within an hour after insertion into obit while in the act of doffing his space suit. This is the earliest appearance of motion sickness in orbital flight on record." In this same individual, while taking 6 dosages of the anti-motion sickness medication scopolamine, he continued to have symptoms until mission day (MD) 5 (Graybiel, 1980). The typical time course was reported to be 1 to 3 days, while "recovery was completed by MD7" (Graybiel, 1975). Two crewmembers observed a delayed onset, after tens of hours, which they attributed to when "stimulus conditions were more stressful". This study was also the first to assess, rather than just note, the provocative role of head movements. One crewmember, on MD 2, executed a standardized series of head movements for a short period, provoking motion sickness, particularly with his eyes open. In a subset of 5 astronauts, the authors (Graybiel, 1980) roughly estimated the number of head movements based upon time and crew activities to attempt to account for the stimuli eliciting motion sickness.

In parallel, Soviet cosmonauts in the Salyut program reported motion sickness "with the aid of special questionnaires" (Matsnev et al., 1983), finding 12 of 27 individuals (44%) presented "some degree" of SMS, while details on severity were not reported. In terms of temporal dynamics, the authors state SMS began "after one-to several hours (or 1-1.5 days) in weightlessness" and lasted "from several hours to 3-7 days" with one individual showing SMS symptoms until MD 14 (Matsnev et al., 1983). Again, the authors noted a "correlation between the development of SMS and motor activity: an increase in the number of head and body movements, especially forward tilts." While details on timing and dosages were not provided, anti-motion sickness medications, along with various non-pharmaceutical countermeasure concepts (e.g., a head cap restricted head on body movements) were used. Finally, 6 crewmembers made two flights and the authors noted, "as a rule during their second flight, the cosmonauts adapted to zero G better; however, they still experienced motion sickness manifestations."

The Space Shuttle program provided the first opportunity for assessing SMS incidence across a larger cohort (Thornton et al., 1987). While there were anecdotal reports of SMS on STS-1 through STS-3, post-flight oral debriefings starting on STS-4 were performed in which SMS symptoms were specifically queried and then graded as mild, moderate, or severe. As the authors note, the examining NASA flight surgeon "usually collected these data during the first hour after landing, and the debriefing required the crewmember to recall his or her experiences during the flight." A series of publications document an increasingly larger cohort (Davis et al., 1988; Davis, Jennings, & Beck, 1993; Thornton et al., 1987), eventually reaching 132 first-time flyers (Davis, Jennings, Beck, et al., 1993), of which 96 reported symptoms of SMS (73% incidence). In terms of severity, 47 (49%) of those were mild, 35 (36%) of those were moderate, and 14 (15%) of those were severe. Including individuals who reported they did not have symptoms (27%), 36% were mild, 26% were moderate, and 11% were severe. In this cohort, there was not a statistically significant difference in SMS incidence between males (86 of 117, 74%) vs. females (10 of 15, 67%), nor an effect of age. A later review (Reschke et al., 2014) of post-flight debriefs reported that female crewmembers who flew on both the Shuttle and ISS tended to have a higher incidence of SMS, though no data or statistics were provided. In an executive summary of sex differences in spaceflight adaptation (Mark et al., 2014), a similar trend was noted but reported as not significant. In an earlier subset of the cohort (Davis et al., 1988), a comparison was made based upon crew position: NASA career astronauts (Commander, Pilot, or Mission Specialist) vs. non-career astronauts (Payload Specialists or Space Flight Participants), finding a trend of higher incidence amongst non-career astronauts (85% vs. 62%), which did not reach significance (p=0.06). Further, 26 crewmembers that had a second flight tended to have a lower incidence (down to 46% vs. 62% on the first flight, in this group), but again this did not reach significance (Davis et al., 1988). However, in these 26 repeat flyers, of the 16 who had SMS on their first flight, 11 also had SMS on their second flight (5 did not). While of the 10 that did not have SMS on their first flight, 9 were also asymptomatic on their second flight (1 had SMS only on their second flight). The mean time between flights was 21 months.

From the postflight oral debriefs, the authors did not provide any data on persistence or temporal dynamics (Davis et al., 1988). On flights after October 1989, there were inflight Private Medical Conferences with NASA flight surgeons and onboard crew medical officers, typically 8-9 hours after main engine cutoff and again 24 hours later, which the authors noted "reduced the reliance of verbal recall postflight" (Davis, Jennings, & Beck, 1993). While data are not shown, the time course of symptoms resolution was reported as 72-96 hours, without taking anti-motion sickness medications (Davis, Jennings, & Beck, 1993). Thornton et al. state temporal dynamics of SMS "onset ranging from minutes to hours, plateaus, and the rapidly resolved in 8-72 hours with 36 hours as average" (Thornton et al., 1987). Finally, a statistical analysis compared astronauts who did not take anti-motion sickness medications to "crewmembers with symptoms of SMS who requested treatment from the crew medical officer. Promethazine was then given by intramuscular (IM) injection in a dose of 25-50 mg adjusted for body weight. A second dose was given 8-10 hours later if symptoms were not resolved." (Davis, Jennings, Beck, et al., 1993). Specifically, the none, mild, moderate, severe ratings given in post-flight debriefings across symptoms of nausea, vomiting, loss of appetite, and stomach awareness were scored into "sick" vs. "not sick" categories. Amongst 20 sick crewmembers treated with IM promethazine on MD1 5 (25%) were still sick on MD2, compared to 74 untreated crewmembers, in which 37 were still sick on MD2 (50%, p=0.046), indicating a more rapid resolution of SMS symptoms with IM promethazine.

Acknowledging the "lack of data at the time the symptoms occur", Reschke et al. (Reschke et al., 2018) recorded verbal reports during and immediately after flight in a single individual who experienced severe spaceflight-associated motion sickness. Specifically, the crewmember was instructed to report motion sickness severity on a scale of 0 to 20, with 20 being frank vomiting and 10 being midway between normal and vomiting. This crewmember vomited over and over on MD1. SMS slowly decreased over days, but even after 7 days was still at a 5 of 20, and had vomited on flight day 5. The crewmember reported that any type of head-on body or whole body movement was provocative. This crewmember was intermittently taking promethazine, either orally or intramuscular injections.

The most rigorous study of SMS was on Spacelab-1, a ten-day Shuttle mission (Oman, 1987; Oman et al., 1986), in which 4 crewmembers (ages 35-53, all male) were thoroughly trained to report motion sickness symptoms prior to flight. The severity and temporal dynamics of SMS were captured inflight using pocket voice recorders and multi-item symptom checklists (Graybiel et al., 1968), an "Overall Discomfort" rating (similar to that later used by (Reschke et al., 2018) in which 10 was to halfway to vomiting), anti-motion sickness medications taken, and sleep periods. Three of the four crew reported persistent symptoms and vomited repeatedly during the first and/or second day of flight. All four eventually used anti-motion sickness medications (0.4mg scopolamine or 25mg promethazine). While prior studies anecdotally identified the provocative effect of head movements, on Spacelab-1, two crewmembers wore head-mounted accelerometers to objectively quantify activity (one of whom provided regular, detailed in-flight motion sickness reports). Further, at the end of the working day on MD 0, 3, and 8, if they were asymptomatic, they were instructed to perform “deliberately provocative head movements”, interleaved with reporting their symptoms. In summary, after several hours of physical activity in orbit, SMS symptoms developed, after which these individuals limited their head movements. Subsequent head movements were associated with further provoking motion sickness, demonstrating the critical role of crewmember motion/activity in SMS (Lackner & DiZio, 2006; Oman, 2002; Oman et al., 1986).

There is less published data on TRMS than for SMS. During capsule water landings with sea state 5 on Skylab 2 (28 days) the crew did not take medications, and one crewmember had severe symptoms, one had mild symptoms, and one had no symptoms. Following this all three crew on Skylab 3 (59 days) took medication (0.35mg scopolamine) 2 hours prior to splashdown, and while the sea state was twice as severe, TRMS was not experienced (Graybiel, 1975; Graybiel et al., 1977), yielding a net incidence of 33% (2 of 6). In the cohort of 27 cosmonauts (Matsnev et al., 1983), 9 individuals reported motion sickness after landing on solid ground (33%), with data presented as pooled across 11 short-duration missions (4, 7, or 14 days) and 5 long-duration (75, 96, 140, 175, 185 days). The large cohort studies (Davis et al., 1988; Davis, Jennings, & Beck, 1993; Davis, Jennings, Beck, et al., 1993) did not report on TRMS, but an earlier paper (Thornton et al., 1987) states that postflight symptoms from Shuttle entry and runway landings were “mild unless deliberately provoked in the early period of re-exposure to gravity.” The largest TRMS study (Bacal et al., 2003) used NASA’s Longitudinal Study of Astronaut Health (LSAH) database to review deidentified medical debriefs that occurred on the first day of return and 3 days later. Following Shuttle missions, 51 of 346 astronaut-flights (where a single astronaut often flew on multiple flights) reported nausea (15%), 8% reported vomiting, and another 3% for dry heaves. Reporting nausea after landing was more common amongst women (30%) than men (13%), which was significant ($p=0.005$), though the authors note concerns about reporting biases (Bacal et al., 2003). Again, in review (Reschke et al., 2014), TRMS was reported more often in post-flight debriefs amongst women vs. men flying on the Shuttle, but more often in men vs. women flying on the ISS (data not provided). The summary notes a slightly higher incidence amongst men, though insignificant (Mark et al., 2014). In the individual case study that experienced severe SMS (Reschke et al., 2018), the authors described the crewmember also having severe TRMS: “motion sickness symptoms during body motion in both pitch and roll persisted for 8 days after the flight”. Beyond this, there is no published, quantified data on TRMS persistence or temporal dynamics.

The incidence of TRMS is positively associated with on-orbit mission duration (Jennings, 1998). Comparing 14 short-duration (mean duration of 7.4 days) Shuttle crewmembers with 18 long-duration (mean duration of 175.8 days) ISS crewmembers, while TRMS was reported by 0% (0 of 14) of the short-duration crew, remarkably 100% (18 of 18) of the long-duration flyers reported being nauseated in interviews 1 to 5 hours after landing on Soyuz (capsule reentry landing on solid ground) (Reschke et al., 2017). Two of the ISS crew experienced such severe TRMS that they did not perform the functional performance tests that were part of this study (Reschke et al., 2017). The authors note that 9 of the combined group who had a history of TRMS took meclizine or promethazine (unspecified) after the deorbit burn. Despite not experiencing TRMS, 7 of these 14 short-duration crew did report being nauseated during the first few days of flight, an indicator of SMS (Reschke et al., 2017). A study of cosmonauts (Gorgiladze & Brianov, 1989), has been cited (Ortega & Harm, 2019; Reschke et al., 2017) as finding that TRMS has an incidence of 27% in short-duration (4-14 days) and 92% long-duration (several months to 1 year) flyers. Further, a book chapter (Ortega & Harm, 2019) reviewing LSAH medical debriefs up through 2013 reports TRMS (referred to as EMS by the authors) incidence across Shuttle and ISS missions. While rare after shorter Shuttle missions of a few days (7-17%, depending upon how possible cases were classified, across Shuttle flights from 1981-86, which tended to be shorter), the incidence appears to increase in later Shuttle flights which were more often 12-14 days (27-56% depending upon possible cases for 1988-02 and 2005-11). Incidence in TRMS was reported to be 47% for ISS missions returning via the Shuttle and 46% when returning via the Soyuz. The authors state that TRMS does not appear to be correlated with age, crew position, or number of previous flights, though data are not presented. In a more recent comparison (Clément et al., 2025) of 10 short-duration (mean duration of 10.4 days) private astronauts to 36 long-duration (mean duration of 208 days) NASA astronauts staying aboard the ISS, TRMS symptoms were queried a few hours after landing on a 0-20 scale. TRMS tended (insignificantly) to be more severe in long-duration flyers (mean score of 9.0/20 vs 4.6/20).

Thus, while several studies have published on SMS and TRMS, there remain notable gaps. First, the vast majority of reports of SMS are made retrospectively in post-landing medical debriefs, where reporting accuracy and precision may be affected by memory and recollection biases. Second, while many studies identify the provoking nature of head movements, only two studies with a few participants (Graybiel, 1975; Oman et al., 1986) attempted to account for head movements in quantifying SMS. In terms of TRMS, head movements have never been accounted for or controlled, nor has persistence been quantified (severity has been assessed in one study (Clément et al., 2025)). Third, the vast majority of published data comes from male, NASA-selected astronauts. In the large-scale studies of SMS incidence (Davis, Jennings, Beck, et al., 1993) only 15 female participants were assessed (compared to 117 males). For TRMS incidence (Bacal et al., 2003), with a similarly imbalanced M/F split of astronaut flights, post-flight nausea was identified to be significantly more common in females than males (noted as not significant elsewhere (Mark et al., 2014)). The opposite trend is present in SMS, warranting further investigation, particularly in a balanced cohort, reporting both SMS and TRMS in real-time. Similarly, one study (Davis et al., 1988) compared SMS incidence in NASA career astronauts with a combined group of spaceflight participants and payload specialists, while another study compared TRMS severity in private astronauts in short-duration missions with NASA astronauts following long-duration missions (Clément et al., 2025).

With the future of commercial spaceflight, it will be critical to quantify SMS and TRMS in a more systematic manner. Here, we report the incidence, severity, and temporal dynamics across a balanced cohort of females and males of commercial flyers, using inflight reporting and standardized head movement stimuli.

Results:

The temporal dynamics of motion sickness, reported using the 0-20 reporting scale following a standardized sequence of head tilts, were captured for each individual crewmember (N=8), both inflight (Figure 1a) and postflight (Figure 1b), with self-reported anti-motion sickness medications shown at the top. Through pre-defined reporting timepoints, it is possible to observe individuals having a wide variety of temporal dynamics of inflight motion sickness. Four crewmembers (S1, 3, 6 and 8) could be characterized as having had a "typical" time course of motion sickness inflight, as has been visualized qualitatively before (Ramsburg et al., 2025; Thornton et al., 1987): motion sickness developed (to varying severities) in the first minutes to few hours inflight, persisted for varying durations (less than a day or ~2 days), before decaying. However, notably, other “atypical” temporal dynamics were also observed in a few crewmembers: 1) motion sickness was very low early in flight before having a delayed onset and increasing after 2-3 days in space (S2 and S7) or 2) after a characteristic early onset and subsequent decay after one day, a relapse in which motion sickness again increased during flight day 3 (S4). These atypical temporal dynamics may have resulted from the temporary protection given from anti-motion sickness medications. Finally, one crewmember (S5) reported no inflight motion sickness across any of their four reports.

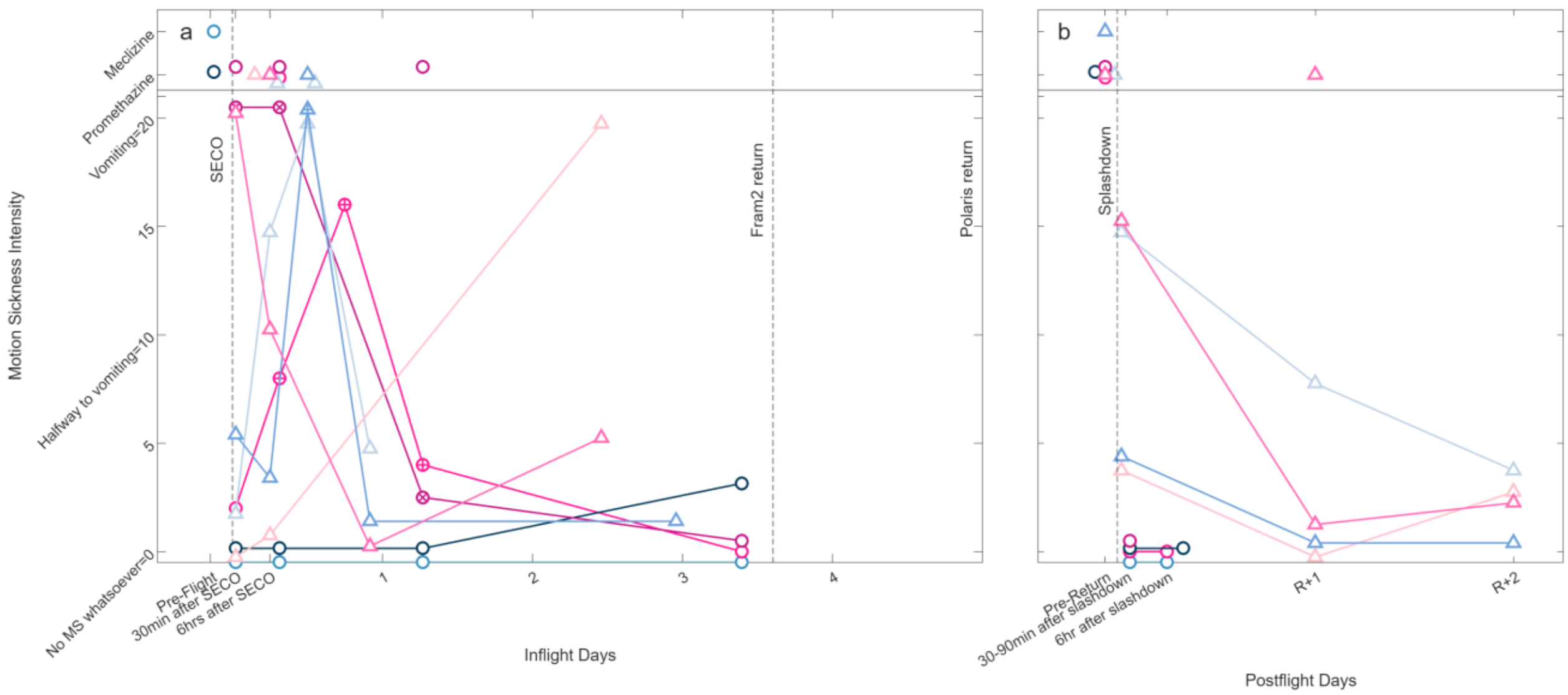

Figure 1: Summary of the intensity of motion sickness over time inflight (panel **a**) and postflight (panel **b**). Individual crewmembers (S1-S8) are shown with unique color shades. Mimicking (Matheson et al., 2026), circles are for Polaris Dawn crew and Fram2 crew are shown with triangles; female crew are shades of pink and males are in shades of blue. When a crewmember felt too motion sick that they chose to not perform the head tilt sequence, the motion sickness report is marked with an ‘x’. Motion sickness reports made separately from the survey and thus without the head tilt sequences are marked with a ‘+’.Vertical jitter has been added to motion sickness intensity reports to avoid symbol overlap. Data are presented similarly to that for a single inflight crewmember on Spacelab in Fig. 5&6 of Oman et al. (Oman et al., 1986), though here we just connect each reporting timepoint linearly, rather than infer a trajectory of motion sickness between timepoints. At the top of the plot, self-reports of anti-motion sickness medications are depicted with corresponding colors/symbols for each crewmember, shown with approximate timing of drug administration, stacked by drug type, and with jitter to avoid overlap. Dosages for both promethazine and meclizine were typically 25 mg, with the following exceptions which were 12.5 mg: S2’s pre-return, S4’s last post-flight, S6’s first inflight. Finally, S8’s second in-flight dose was 50 mg, while their third and pre-return dose were 12.5 mg.

Postflight, none of the Polaris Dawn crew reported motion sickness. In fact, these crewmembers included comments in their responses to the open-ended survey response emphasizing the stark lack of TRMS: “I'm very prone to motion sickness but experienced no symptoms”, “No feelings of sickness, head feels very heavy though”, “I don't think I am experiencing motion sickness, but I am having an interesting time adapting to gravity. Occasionally I lose my center of balance, the world moves in a way I don't expect" and "None". The Fram2 crew each experienced TRMS shortly after flight (to varying degrees), which decayed over two days, though on the last report approximately 48 hours after splashdown 3 crewmembers still reported slight motion sickness.

Use of anti-motion sickness medications was common, with 6 crewmembers reporting inflight use (and the other 2 pre-flight) and 6 taking medications just prior to return. Promethazine (typically 25 mg) was most commonly used, though meclizine (typically 25 mg) was also used by two crewmembers.

Incidence

To align with prior work that has focused on the incidence (typically defined as experiencing at least slight motion sickness) (Davis et al., 1988; Davis, Jennings, & Beck, 1993; Davis, Jennings, Beck, et al., 1993; Thornton et al., 1987), here we compute whether each crewmember reported >0 at any time point, inflight as well as postflight (Figure 2). Inflight 7 of 8 crewmember reported at least some motion sickness (incidence = 87.5%, 95% confidence interval (CI) for proportion = 47.35 to 99.68%), while postflight only 4 of 8 reported (incidence = 50.0%, CI = 15.7 to 84.3%). Since each crewmember made reports both inflight and postflight, we were able to quantify those that experienced SMS vs. TRMS (Figure 2b). Half (4 of 8) reported both SMS and TRMS. However, 3 crew experienced SMS, but not TRMS. Finally, one individual did not report SMS nor TRMS, while no one experienced TRMS postflight without inflight SMS. Leveraging the within participant reporting inflight and postflight, there was an insignificant trend of less incidence of TRMS than SMS (McNemar's exact test, $\chi^2$=1.33, p=0.25)

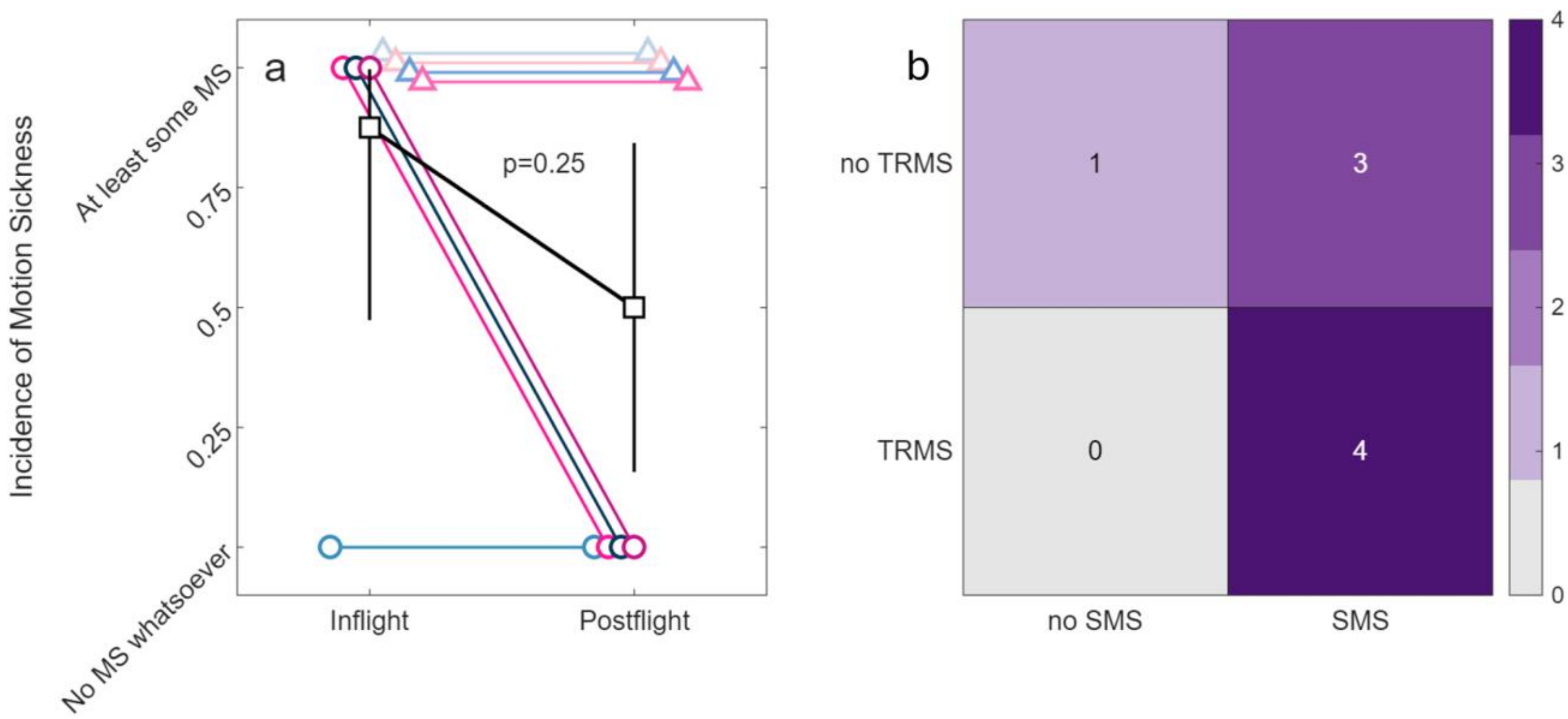

Figure 2: Incidence of motion sickness inflight vs. postflight. Panel **a** contrasts incidence inflight to that postflight. Horizontal and/or vertical jitter were added to avoid symbols and lines overlapping. Incidence rate with 95% confidence intervals of the binomial distribution are shown. There was an insignificant trend for lower incidence of motion sickness postflight (50%) compared to inflight (87.5%). Panel **b** shows the contingency table of inflight (SMS) and postflight (TRMS) motion sickness.

While incidence is typically defined as experiencing at least slight motion sickness, we can also report incidence of vomiting. Inflight, 5 of 8 crewmembers (1 on Polaris Dawn and all 4 crewmembers on Fram2) reported vomiting (62.5% incidence, CI: 24.5 to 91.5%), while no crew reported vomiting postflight after these short-duration flights (0%, CI: 0 to 36.94%).

Severity

From the multiple reports of motion sickness, a "severity" metric was extracted, capturing the most severe report, inflight and postflight, for each crewmember (Figure 3). Notably, inflight motion sickness was quite severe for several crewmembers, with five (62.5%) reporting 20 (i.e., vomiting). Following these short duration flights, postflight motion sickness severity was significantly lower than that inflight (paired t-test, t(7) = 3.76, p=0.007, averaging 14.9 vs. 4.8 of 20, with mean SMS vs. TRMS difference of 10.1, 95% CI = 3.8 to 16.5). While two subjects had severe postflight motion sickness (15 of 20), two were much more mild (4 of 20), and again, all four Polaris Dawn crew did not experience any TRMS.

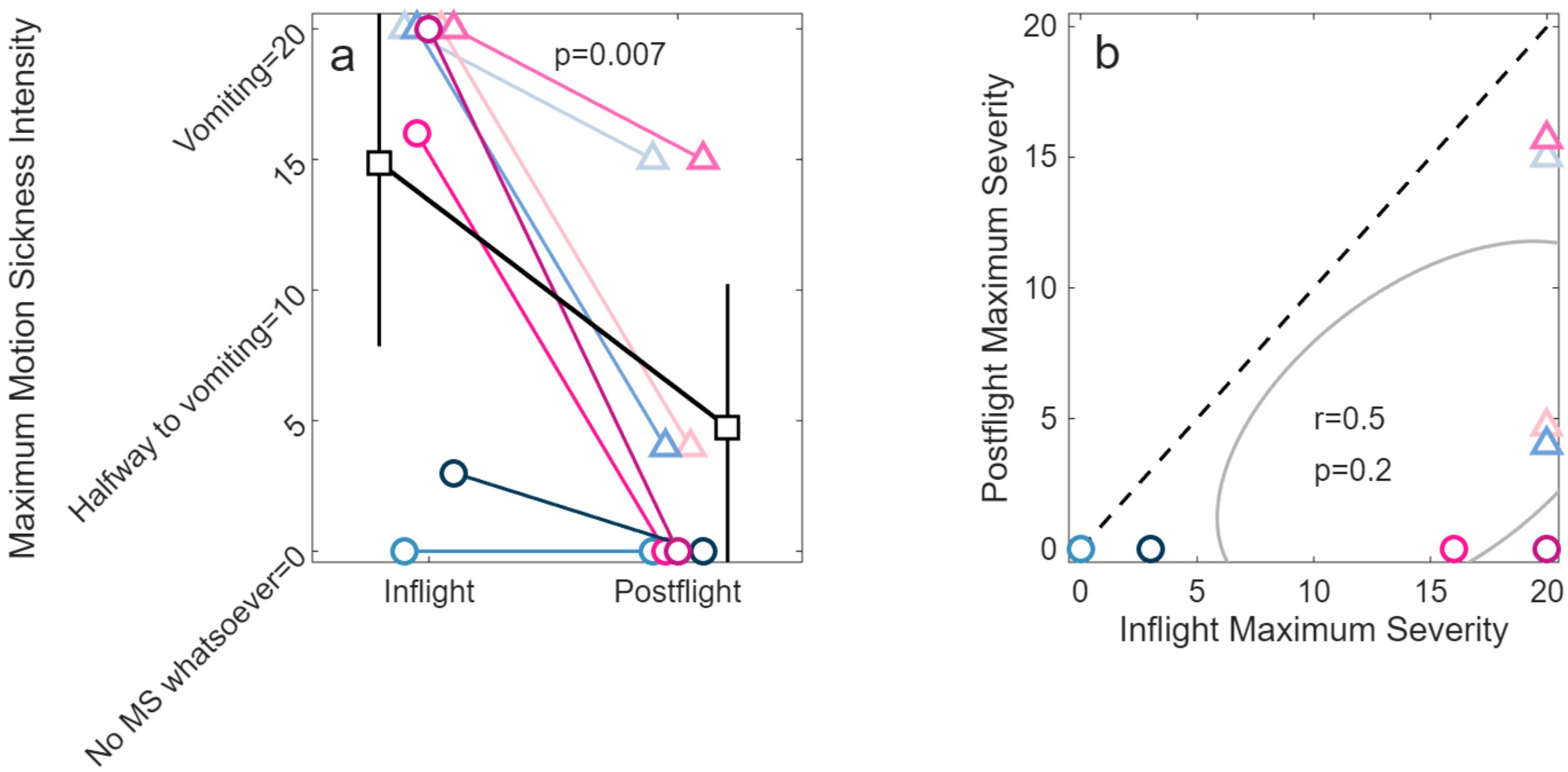

Figure 3: Maximum severity of motion sickness inflight vs. postflight. Panel **a** contrasts each crewmember's inflight maximum severity to that postflight. Horizontal jitter was added to avoid symbol overlapping. To be consistent with Figure 2, means and 95% confidence intervals are shown. There was statistically significantly lower peak severity of motion sickness postflight compared to inflight. Panel **b** relates each crewmember's inflight maximum severity to that postflight. Vertical jitter was added to avoid sets of triangles overlapping. The dotted unit line represents equivalent maximum severity postflight as inflight. A correlation ellipse shows a slight positive, but insignificant, correlation between inflight and postflight maximum severity of motion sickness experienced by individual.

Figure 3b depicts the association between each crewmember's severity of SMS and TRMS. All symbols being below or on the unity line highlights that no crewmembers reported more severe motion sickness postflight than that reported inflight. To assess whether crew that experienced more severe motion sickness inflight also had more severe reports postflight, we tested for a Pearson correlation which found an insignificant, positive trend (r=0.5, p=0.2). As several reports were at 0 or 20 (stressing the normality assumption), we also report analysis using a Spearman-rank correlation and found a similar outcome (r=0.71, p=0.14).

Persistence

Finally, we extracted a metric of persistence from the temporal dynamics captured in Figure 1 (timing of the first report for which the motion sickness rating remained <3, see Methods: Data processing and statistical analysis, for details). With this definition of persistence, inflight motion sickness continued on average for 1.85 days (or 44.4 hours, 95% CI: 0.81 to 2.9 days). Three crewmembers had their last inflight report remain above 3, such that this timepoint was used as a proxy for the persistence metric, while their motion sickness may have remained even longer into their flight. Postflight motion sickness persisted for an average of 0.78 days (18.7 hours, 95% CI: 0.065 to 1.50 days), which was slightly (but insignificantly) shorter than inflight (paired t-test, $t(7)=1.89$, $p=0.1003$, 95% CI: -0.27 to 2.41 days). Figure 4b depicts the relationship between each individual's persistence in motion sickness inflight vs. Postflight.

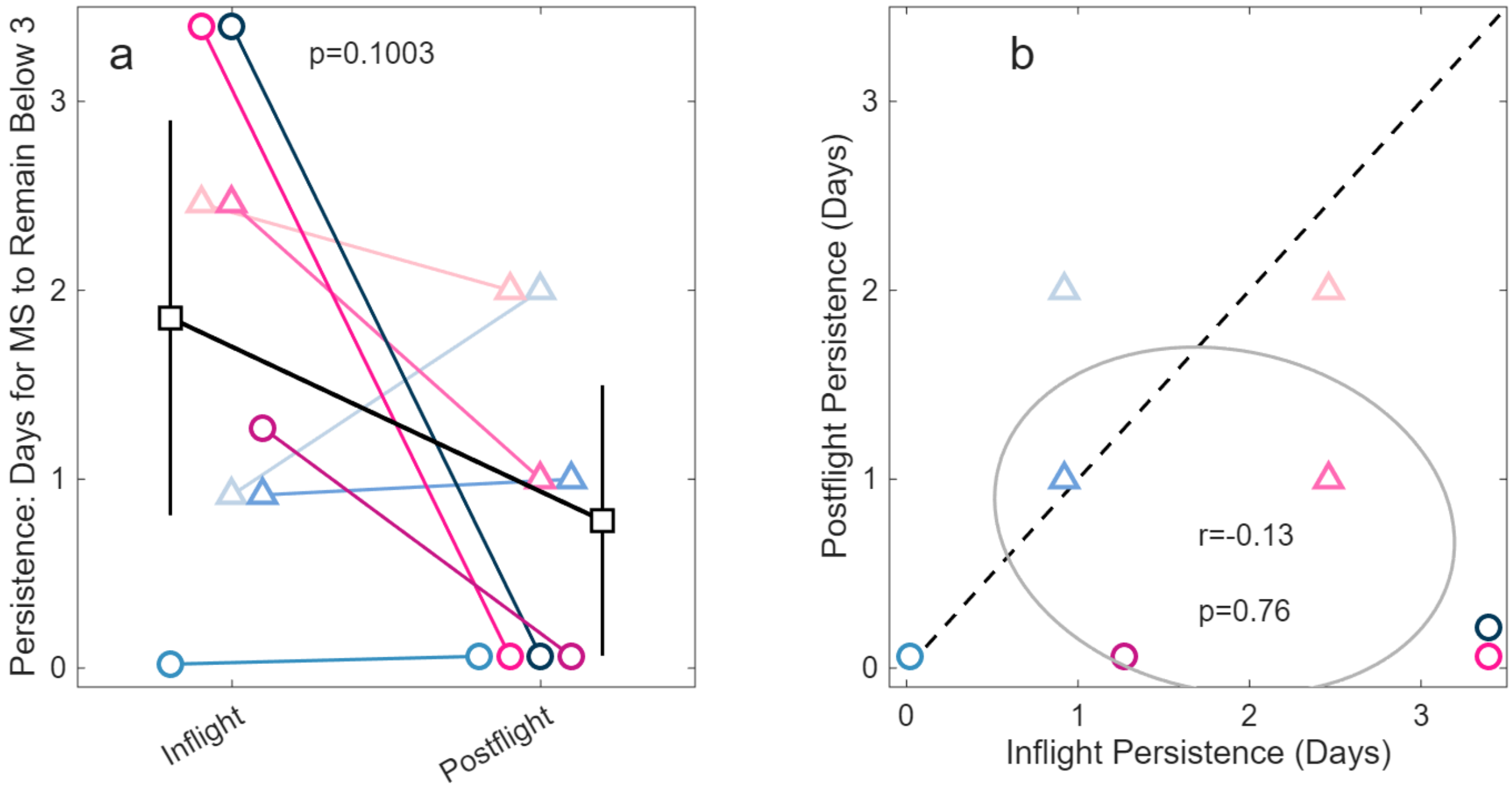

Figure 4: Persistence of motion sickness inflight vs. postflight. Persistence was quantified as the number of days in each mission phase (inflight or postflight) until motion sickness reports dropped below (and remained below) three. Panel **a** contrasts each crewmember's inflight persistence of motion sickness to that postflight. Horizontal jitter was added to avoid symbol overlapping. There was an insignificant trend for shorter motion sickness persistence postflight. Panel **b** relates each crewmember's inflight motion sickness persistence in days to that postflight. Vertical jitter was added to avoid the rightmost circles overlapping. The dotted unit line represents equivalent persistence of motion sickness postflight as inflight. A correlation ellipse is shown capturing an insignificant relationship between inflight and postflight persistence of motion sickness.

Additional Analyses

As a secondary focus, we explored correlations between individual attributes and severity of inflight and postflight motion sickness. First, in this cohort, severity did not differ between males and females, for either inflight SMS ($p=0.18$) or postflight TRMS ($p=1$). Next, there was a trend towards less motion sickness experienced during Polaris Dawn compared to Fram2 missions inflight ($t(6)=2.1$, $p=0.08$) which was significant postflight (t(6)=2.99, p=0.024). There were no correlations with crewmember age, either inflight (p=0.82) or postflight (0.49). Similarly, crewmember MSSQ (self-report questionnaire of history of terrestrial motion sickness) score did not correlate with inflight (p=0.88) or postflight (p=0.46) motion sickness severity (same conclusion if using MSSQ percentile rather than score).

Discussion:

In this article, we provide a comprehensive analysis of spaceflight-associated motion sickness (both SMS and TRMS) in short-duration commercial astronauts of the Polaris Dawn and Fram2 missions aboard the SpaceX Crew Dragon spacecraft. Historically, motion sickness reports have been limited to agency (e.g., astronaut or cosmonaut) populations, which undergo rigorous physiological selection (though motion sickness susceptibility is not part of this criteria). This can lead to population biases in health outcomes risking generalizability to other populations (reviewed in (Gray & Johnston, 2008)). Moreover, spaceflight-associated motion sickness data recorded to date have historically come from predominantly male astronauts. In this article, we evaluate the incidence, severity, and time course of both SMS (here, motion sickness following transition from Earth gravity to microgravity) and TRMS (motion sickness following the reverse transition from microgravity to Earth gravity) in eight commercial flyers, evaluating an equal distribution of male and female participants.

Of our eight participants, almost all (7/8; 87.5% incidence) reported a degree of SMS, while only half (four; 50.0% incidence) reported some degree of TRMS. Regarding SMS, over half of the participants reported vomiting in orbit (5/8; 62.5% vomiting incidence), with reports of vomiting subsiding following FD 3. Our results contrast many grouped reports of agency incidence rates, which we note are inconsistent with one another and depend greatly on which programs were included. As outlined in the introduction, NASA incidence reports have demonstrated an increase over time: the Mercury and Gemini astronauts reported 0% incidence of motion sickness; Apollo astronauts reported 36% incidence; Skylab astronauts reported 56% incidence; and in a large comprehensive Shuttle investigation, astronauts reported 73% incidence. Previous studies (Graybiel, 1980; Reschke et al., 1998) have suggested increasing habitat volumes contribute to increased SMS, though others note that head movement activity within the volume is more likely the causal stimulus (Lackner & DiZio, 2006; Oman, 1987; Oman et al., 1986). All crew in our cohort flew in SpaceX's Crew Dragon with a habitable volume of 9.3 $m^3$ for 4 crew, which is larger than Mercury (1-1.7 $m^3$), Gemini (~2.4$m^3$ for 2 crew), and somewhat comparable to the Apollo Command Module (~6$m^3$ for 3 crew), while smaller than the Space Shuttle Orbiter (~70 $m^3$) (Nicogossian & Uri, 1994). Depending on which cohorts were included (and criteria used), pooled agency motion sickness incidence has been described as approximately 40% (Putcha et al., 2012), 50% (Thornton & Bonato, 2013), 60% (Ortega et al., 2019), and ranging from 60-80% (Heer & Paloski, 2006). A contributing factor in this variation is the substantial variation in susceptibility to spaceflight associated motion sickness. In our dataset, inflight and postflight severity ranged the full scale from 0 (no motion sickness whatsoever) to 20 (vomiting). Previous work has explored various attempts of predicting pre-flight inter-individual differences (Lackner & DiZio, 2006; Oman, 1987; Oman et al., 1986) that are often stated to be unreliable (Graybiel, 1980) likely due to operational confounds in the provocative stimulation (Oman, 1990), though some pre-flight tests querying vestibular function (Diamond et al., 1990; Diamond & Markham, 1991, 1992) and adaptive responses show promise (Bles et al., 1996, 1997; Nooij et al., 2007; Ockels et al., 1990). Predicting inter-individual differences was outside the scope of our study, so it is unsurprising with our small sample size that there were not significant correlations for SMS or TRMS with age or MSSQ (Earth-based reported motion sickness history), as these were instead aimed at characterizing this cohort.

It is interesting the Polaris Dawn crew did not report TRMS, while all Fram2 crewmembers did. This may have been related to the motion experienced by the two capsules after splashdown (Table 1). Inertial measurement unit metrics for capsule motion (the magnitude of the body-rate RMS and peak, specific-force RMS, and peak-to-peak hang angle) were all roughly twice as large during Fram2 compared to Polaris Dawn. While quantifying the capsule motion provides relevant insight into the provocative motion simulation for TRMS, these capsule motions do not fully capture each crewmember's motion stimulation based upon seat position and any head movements made. The Fram2 crew also participated in a separate experiment in which they performed self-egress from the capsule on the recovery ship, rather than an aided egress, which could have further provoked motion sickness. Egress on the ship happened after the first post-splashdown report of motion sickness, but before the report at "6 hr after splashdown". While TRMS may have been impacted by differences in sea state between the two missions, the trend of more severe reports of SMS by the Fram2 crew were presumably not due to differences in environmental stimuli (both on orbit, in microgravity, in same capsule).

Caution should be used in comparing our results to prior investigations' incidence rates that used different methodologies, particularly those obtained without a provocative stimulus prior to querying motion sickness. Various scales for reporting severity (e.g., none/mild/moderate/severe (Davis et al., 1988) vs. 0-20 anchored by vomiting (Clément et al., 2025; Oman, 1987; Oman et al., 1986; Reschke et al., 2018)) further complicate comparisons. Historical agency motion sickness reports for SMS may also be confounded by post-hoc recollections and by agency-level pressures. For example, there may be difficulty recalling SMS incidence, and symptom reporting could be impacted by pressures to perform and contribute to current or upcoming missions. Therefore, incidence and severity may have been systematically underestimated in some agency datasets, with this being less of a concern with real-time reporting in crew decoupled from future flight assignments. A potential advantage of our study methods is the first implementation of discreet, real-time reporting into a digital questionnaire, which may help minimize reporting biases.

Due to these combined factors, it is difficult to directly compare our results to most prior agency astronauts' reports; rather, our goal was to more systematically and reliably assess motion sickness, providing a baseline evaluation for future studies. While the incidence and severity of SMS in this cohort tend to be higher than agency fliers, they closely mirror those reported during the NASA Spacelab-1 mission (Oman, 1987; Oman et al., 1986). This mission notably also used stimulating head motions prior to querying SMS symptoms reported in real-time. We acknowledge that in the operational environment of spaceflight, head movements between the standardized provocative sequence were uncontrolled and differences in crew activities are likely to contribute to inter-individual differences in spaceflight associated motion sickness. Further, some motion sickness reports were provided separately from our pre-planned tablet questionnaires and without the head tilt sequence, which we integrated into our dataset and noted with '+' symbols in Figure 1. Integrating both pre-planned reports with the head tilt sequence with ad hoc reports provides a more complete summary of spaceflight associated motion sickness, but we acknowledge that how often reports are provided can bias metrics of incidence, severity, and persistence. In future missions, crew activity also will not be restricted, and functionally the provocative head tilt sequence may be analogous to the motion sickness induced by an acute activity required during spaceflight operations (e.g., doffing a launch suit).

The lower degree of TRMS vs. SMS for short-duration flights is consistent with historical reports, but we find a greater incidence and severity of TRMS in commercial flyers compared to agency counterparts, reported following Shuttle landings. Despite the two commercial missions being relatively short in duration (3.6 and 4.9 days in microgravity), TRMS incidence here was substantially higher (50%, CI: 15.7-84.3%) than has been reported in longer, short-duration (7.4-day mean duration flights) agency flyers, of 0% across 14 crewmembers in one study (Reschke et al., 2017). However, TRMS reported here may have been exacerbated by the sea-state of splashdown (vs Shuttle landings with lower entry g's from which these agency reports were derived). In capsule water landings during Skylab (Graybiel, 1975; Graybiel et al., 1977), 2 of 6 crew (33%) reported TRMS. Here our maximum severity of TRMS averaged 4.8 (CI: -0.7 to 10.2) out of 20, while in 10 longer, short-duration (mean of 10.4 days) private astronauts using the same scale (though without standardized head movements and only queried at a single time point a few hours after landing) had a comparable mean of 4.6 (Clément et al., 2025). Future work should disambiguate the relative contributions towards TRMS, between the transition from microgravity back to Earth gravity vs. simply seasickness from capsule motion without external visual references, or an exacerbating effect of sea state provoking sensory conflict in microgravity-adapted crew. Because TRMS incidence and severity have historically been higher for long-duration flights in astronaut populations (100% TRMS incidence for 175.8-day mean duration flights in 18 agency ISS crew (Reschke et al., 2017), and mean TRMS severity of 9.0 of 20 in 36 crew averaging 208 days in space (Clément et al., 2025)), TRMS incidence and severity can be expected to increase from the results presented here as mission durations extend for commercial flyers.

Our results also provide new insight for female flyers, which have been historically underexamined. While our sample is small and should be built upon in future investigations, we found a 100% incidence of SMS among these four female commercial flyers, and a 50% incidence of TRMS. Notably, three of four female crewmembers experienced inflight vomiting and all four took anti-motion sickness medications inflight. While comparisons should be cautioned given the sample samples and differing methodologies, we note that one study found no significant differences in SMS in agency flyers (Davis, Jennings, Beck, et al., 1993), while other studies have conflicting reports (Mark et al., 2014; Reschke et al., 2014). With only one participant in the current cohort, we cannot analyze the effect of previous flight experience without identification, but note that prior studies found this reduces SMS incidence (Davis et al., 1988). Future work should disentangle the contributions of prior neurovestibular adaptations to gravity transitions vs. altered head movement activities learned from prior spaceflight experience.

Another underexamined aspect of SMS and TRMS in both agency and commercial flyers is the persistence (i.e., sustained duration) of motion sickness symptoms, as well as temporal dynamics. Here we quantify both, finding that SMS persisted for at least two days in half of flyers (4/8; 50%), and at least three days in a quarter of flyers (2/8; 25%). In one individual, vomiting persisted until at least the third day on orbit despite taking promethazine on FD 1. In contrast, TRMS persistence trended lower than SMS persistence (while not significant), with only a quarter of flyers having TRMS persist for at least two days. Furthermore, SMS persistence was negatively correlated with TRMS persistence (also not significant). In fact, the two flyers (1M and 1F) with the most persistent SMS reported no TRMS. We acknowledge that persistence (as well as incidence and severity) may be biased due to the uneven number of inflight reports (3-5) vs. postflight reports (1-3), as well as the standardized timings vs. additional reporting, where some crew did not continue to report until symptom resolution (requiring a proxy for the persistence metric) and also could have not reported a brief period of motion sickness.

Further, some crew provided their first inflight “30 min after SECO” report earlier (roughly after 5 min for S5, 12 min for S6, 23 min for S7 and S8), at which point symptoms may have not had time to develop, impacting measures of symptom onset. However, with the intended standardized timing of reports, rather than only ad hoc or post-flight reports, we were able to quantitatively capture the temporal dynamics of motion sickness onset, plateau, and recovery, with four participants consistent with this “typical” (Ramsburg et al., 2025) prior time course of SMS (without data) of onset within hours to minutes and resolution within 8-72 hours with 36 hours as an average (Thornton et al., 1987) or 72-96 hours (Davis, Jennings, Beck, et al., 1993) or separately reported of a typical time course was reported to be 1 to 3 days, while “recovery was completed by MD7” (Graybiel, 1975), with some isolated reports of as long as 14 days (Matsnev et al., 1983; Reschke et al., 2018). Persistence longer than about 3 days could not be captured with the timing of reports in our study combined with the shorter flight duration, but it is notable that symptoms did not always resolve by the last report inflight or postflight. We were able to capture atypical temporal dynamics in some individuals that warrant further investigation with standardized timing of reports. Anti-motion sickness medications may contribute to delayed onset or relapse in symptoms, though given our small sample size and non-randomized nor placebo-controlled drug administration we cannot assess the efficacy. Figure 1 simply presents the medications optionally reported by each crewmember. Future work should aim to systematically assess the efficacy of different medications (promethazine (Davis, Jennings, Beck, et al., 1993), meclizine, scopolamine, etc. vs. placebo), assessing dosages and timing of administration, for mitigating spaceflight SMS and TRMS (Buckey et al., 2026), to build upon double-blinded, ground-based studies (Weerts et al., 2014). Our time course of decay in TRMS symptoms is the first quantification of temporal dynamics of motion sickness postflight.

Consistent with these results, the theoretical relationship between SMS and TRMS is not straightforward and likely depends on numerous factors including flight duration and individual differences in rate of (re)adaptation to altered gravity environments. According to the vestibular sensory reinterpretation hypothesis, motion sickness abates following central adaptation to the microgravity environment, once altered patterns of vestibular sensory information are correctly expected by the central nervous system (Oman, 1998; Young et al., 1984b, 1986). Similarly, TRMS would be expected to abate following readaptation to the Earth gravity environment and correct central expectations of the vestibular sensory information produced within it. For individuals that do not fully adapt to the microgravity environment (particularly likely during short-duration flights), motion sickness can be expected to persist during flight (resulting in sustained SMS); however, it’s possible that a partially adapted central nervous system may be more readily situated for Earth gravity following return to Earth (resulting in less severe and less persistent TRMS). Similarly, an individual that rapidly adapts to microgravity may have less persistent or severe SMS, but be more vulnerable to TRMS upon return to Earth. Conversely, individuals may be general ‘fast adapters’ and be protected from both transitions, or otherwise be ‘fast readapters’ and be less susceptible to TRMS upon return to Earth (see (Allred et al., 2023) for a theoretical modelling overview of individual factors influencing sensory conflict and adaptation rate to gravity transitions). For long-duration flights, with complete neurovestibular adaptation to microgravity, a clearer, positive correlation between an individual’s SMS and TRMS severity and persistence may be anticipated.

Future work should capture the temporal dynamics of SMS and TRMS with comparable methods to examine intra-individual trends in long-duration flights. NASA currently reports administering motion sickness questionnaires regarding the first three days of flight as part of the Spaceflight Standard

Measures program (Hardy et al., 2025), which may not extend long enough to fully capture SMS persistence. Quantifying the incidence, severity, and time course of SMS and TRMS, in agency and commercial crews, can help inform mission operations and development of pharmaceutical (Buckey et al., 2026; Davis, Jennings, & Beck, 1993; Weerts et al., 2014) or non-pharmaceutical countermeasures (Lonner et al., 2023, 2025).

## Materials and Methods:

### Subjects and pre-flight assessments:

The Polaris Dawn and Fram2 crewmembers participated in this study (4M/4F, ages 28-62 years). One crewmember had previously flown in space for approximately three days, while the other seven were first time flyers. The protocol was reviewed by the University of Colorado Institutional Review Board. All subjects agreed to participate at an informed consent briefing.

Prior to their missions, each crewmember completed the Short-Form Motion Sickness Susceptibility Questionnaire (MSSQ-Short). The MSSQ (Golding, 1998, 2006) has commonly been used to assess an individual's susceptibility to motion sickness, including in spaceflight applications (Bretl & Clark, 2020; Lonner et al., 2023), by asking about a history of experiences with terrestrial forms of motion sickness. MSSQ scores were 2, 3, 5.95, 10.29, 15.75, 16, 20, and 30, ordered from lowest to highest, which corresponds to the 10$^{th}$, 15$^{th}$, 28$^{th}$, 46$^{th}$, 65$^{th}$, 66$^{th}$, 77$^{th}$, and 94$^{th}$ percentiles (mean of 50.1$^{st}$ percentile, such that this cohort was of average susceptibility to motion sickness on Earth, as expected from a random sample).

### Orbital spaceflight mission architectures:

The Polaris Dawn crew (shown as circles throughout) spent nearly five days (4.9 days; 118 hr, 13 min) in orbit. The Fram2 crew (shown as triangles throughout) spent just over three and a half days (3.6 days; 86 hr, 33 min) in orbit. Thus, both missions are best characterized as short-duration flights. Further, both missions occurred aboard the SpaceX Crew Dragon spacecraft, which has a habitable volume of approximately 9.3 $m^3$.

On Polaris Dawn (Menon et al., 2025), all four crew members participated in an extravehicular activity (EVA), with two exiting the vehicle while two crew members remained in the Dragon cabin while being exposed to space (Matheson et al., 2026). For both missions, launch ascent on the Falcon 9 rocket provided g-forces peaking at approximately 4 g's primarily in the sagittal, x-axis direction (i.e., "eyeballs in") for approximately 8-9 minutes until second stage engine cutoff. Following microgravity coasts in orbit (durations detailed above), accelerated re-entry was initiated. Re-entry from the deorbit burn to splashdown took approximately 40 minutes, with g-forces peaking also at approximately 4 g's primarily in the same x-axis direction. Both splashdowns occurred during mild sea state conditions off the coast of Florida for Polaris Dawn and off the coast of California for Fram2, with the capsule hoisted onto the recovery ship within about 35 minutes following splashdown.

Table 1 shows inertial measurement unit metrics of each capsule's motion computed over the period from 2 minutes to 20 minutes after splashdown. While the hang-angle (capsule tilt relative to vertical) and the dominant period of the wave rocking were similar, the magnitude of the body-rate RMS and peak, specific-force RMS, and peak-to-peak hang angle were all roughly twice as large during Fram2 compared to Polaris Dawn. Crewmembers were egressed out of the capsule just under an hour after splashdown.

Table 1: Metrics of capsule motion post-splashdown for Polaris Dawn vs. Fram2

| IMU metric* | Units | Polaris Dawn | Fram2 |
|---|---|---|---|
| Body-rate magnitude RMS | deg/s | 3.7 | 8.0 |
| Body-rate magnitude peak | deg/s | 10.2 | 21.6 |
| Detrended specific-force RMS | $m/s^2$ | 0.76 | 1.55 |
| Hang-angle mean | deg | 17.0 | 18.8 |
| Hang-angle peak-to-peak | deg | 17.7 | 30.6 |
| Dominant period (body-axis-1 accel) | s | 6.7 | 6.3 |

* The body-rate magnitude RMS is the root-mean-square error of the Euclidean magnitude of the three-axis angular-rate vector, measuring how hard the capsule rotated (rocking/yawing) on the water. The body-rate magnitude peak is the same measurement, but rather than the RMS over the 18-minute window, this is the single-sample peak, so is more sensitive to a brief jolt (wave, boat wake, recovery contact) than RMS. The detrended specific-force RMS the root-mean-square error of the magnitude of the three-axis specific-force vector minus its mean over the 18-minute window. Subtracting the mean aims to remove the steady gravity/tilt bias, so this metric reflects the oscillating acceleration. The hang angle is the capsule tilt relative to vertical while on the water. The mean hang-angle captures how far over the capsule typically sits and similar means reflect similar average capture tilt, not similar wave energy. The peak-to-peak hang angle is the maximum hang angle within the 18-minute window minus the minimum hang angle, capturing how far the capsule swings in tilt. The dominant period is computed by the strongest spectral peak of detrended specific-force in the 0.02-0.5 Hz band (swell / capsule-on-water band). Similar periods imply both capsules rocked at about the same cadence.

Crewmembers performed numerous operational and scientific activities in parallel to this study that may have influenced their motion sickness reports that could not be controlled for. While standardized head movements were performed immediately before querying motion sickness symptoms (as detailed in the following subsection), individual crewmembers each experienced unique motions while on orbit and post splashdown. These activities may have influenced their motion sickness symptoms both directly via the additional provocative motions, but also indirectly by modifying the extent of their individual adaptation to the transitioned-to gravity environment.

Inflight and postflight study protocols:

As noted, most prior investigations that aimed to quantify motion sickness experienced during and shortly after flight have not aimed to control for motion-related activity or exposure, despite the evidence that head movements exacerbate spaceflight-associated motion sickness (Lackner & DiZio, 2006; Oman et al., 1986). Here, to provide a standardized provocative stimulus immediately before crew report motion sickness symptoms, crewmembers performed a specified series of head-on-body tilts: first tilting their head in roll with the left ear to the left shoulder, then right ear to right shoulder, then back up to upright, then pitch forward nose towards chest, then backward with nose up, with the full sequence repeated four times. This series of head movements was motivated by prior studies with a few astronauts who performed a similar sequence (Graybiel, 1975, 1980; Oman, 1987; Oman et al., 1986). Crewmembers were instructed to move their heads smoothly, such that going from one position to another would take approximately one second. The magnitude of the head tilt was instructed to be “tilting your head to the limit which is comfortable, without straining”. This was estimated to be +/-30 degrees in roll left and right, 70 degrees in pitch forward, and 50 degrees in pitch back. Crewmembers were trained in performing the head tilt sequence, and during practice pre-flight, were given feedback from test operators (e.g., “you are moving too quickly”). Inflight and postflight, crewmembers were re-provided these instructions regarding the head tilt protocol. If crewmembers felt too motion sick, they were given the option to abstain from the head movements, but to still report their motion sickness level. Some crewmembers opted for skipping the head movements, each of which are noted with an ‘x’ in Figure 1.

Motion sickness intensity ratings and medications:

Following the head tilt sequence, crewmembers reported motion sickness intensity using a simple 0 to 20 whole number scale, where 0 corresponded to “no motion sickness whatsoever”, 20 corresponds to “vomiting or immediately about to vomit”, 10 corresponded to “halfway to vomiting”. This reporting of intensity has been used extensively to quickly assess the intensity of terrestrial motion sickness (Bretl et al., 2019; Bretl & Clark, 2020; Cheung et al., 2007; Jarchow & Young, 2007; Young et al., 2001) and correlates well with more complex scales (Hecht et al., 2002). The reporting scale was similar to that used in a few previous spaceflight studies (Clément et al., 2025; Oman, 1987; Oman et al., 1986; Reschke et al., 2018) and provides more resolution than none/mild/moderate/severe distinctions (Davis et al., 1988; Davis, Jennings, & Beck, 1993, 1993), without the details of specific symptoms (Graybiel, 1980; Graybiel et al., 1968). Crewmembers reported their motion sickness intensity discreetly through a Qualtrics-style survey on a tablet. After reporting intensity, crewmembers optionally reported any other comments or context, including whether they had used any anti-motion sickness medications, and if so, when, what medications, and dosages. Each report was nominally made at the time to avoid the impact of recollection. In a few cases, responses were entered into the tablet hours later, but reports of inflight motion sickness experiences were always made while inflight, rather than in postflight debriefs (Bacal et al., 2003; Davis, Jennings, & Beck, 1993; Jennings, 1998; Mark et al., 2014; Reschke et al., 2017). Digital survey reports were reviewed by the authors familiar with the mission to identify and correct any erroneous or missing data. One crewmember entered ‘2’ during a bout of vomiting (confirmed by the crewmember in a comment for a subsequent report), which was deemed an entry error and corrected to ‘20’. In another instance, a crewmember was too motion sick (i.e., vomiting) to complete the survey, so an entry was added with a corresponding ‘20’. In some cases, crew verbally reported over space-to-ground communications regarding motion sickness experienced apart from the pre-defined time points in which they made head movements and completed tablet questionnaires. The same 0-20 reporting scale was used and these data were integrated into the analysis but denoted with + filled symbols. Similarly, medications that were administered, but not captured in questionnaires were integrated into the analysis.

Time course of measurements:

To quantify the time course of motion sickness intensity, crewmembers completed the sequence of head tilts and reporting at various time points throughout their mission (in-flight at approximately 30 minutes and 6 hours after second engine cutoff (SECO) and then daily afterwards, post-flight at approximately 30 minutes and 6 hours after splashdown and then daily afterwards for two days). Due to operational constraints, some planned time points of head tilts and motion sickness reporting were skipped or shifted slightly (see Figure 1 for time points which were completed for each crewmember).

Data processing and statistical analysis:

In addition to capturing the individual crewmembers' motion sickness intensity over time, we computed metrics for incidence, severity, and persistence. As has been noted previously (Thornton et al., 1987), "incidence depends upon the observer's criteria." As the most comprehensive datasets (Davis, Jennings, Beck, et al., 1993), consider even "mild" categorization of motion sickness as being a positive case, we consider any report greater than 0 (more than "no motion sickness whatsoever") as being a positive case (also report on incidence of vomiting). For severity, the maximum reported motion sickness intensity reported by each crewmember was computed from the 3-4 inflight reports made (Inflight) and then again amongst the 1-3 postflight reports made (Postflight). For persistence, we computed the timing of the first report for which the motion sickness rating remained <3, from SECO for inflight and from splashdown for postflight. Inflight, 3 of 8 crewmembers had their last report at or above 3, so we used the time of their last report as a proxy, such that it was known motion sickness persisted for "at least" this duration. Alternative criteria for persistence are covered in the Discussion.

We performed within-subject statistical analyses contrasting inflight to postflight for each of incidence (McNemar's test), severity (paired t-test), and persistence (paired t-test). Next, we assessed associations of an individual's inflight outcomes with their reports postflight, for severity and persistence (Pearson correlations, as well as Spearman Rank due to the potential for violations of normality). Prior to applying parametric statistics, we visualized the data for substantial deviations from normality assumptions.

References

Allred, A. R., Kravets, V. G., Ahmed, N., & Clark, T. K. (2023). Modeling orientation perception adaptation to altered gravity environments with memory of past sensorimotor states. *Frontiers in Neural Circuits*, *17*. https://doi.org/10.3389/fncir.2023.1190582

Bacal, K., Billica, R., & Bishop, S. (2003). Neurovestibular symptoms following space flight. *Journal of Vestibular Research*, *13*(2–3), 93–102. https://doi.org/10.3233/VES-2003-132-304

Berry, C. A., & Homick, G. L. (1973). Findings on American Astronauts Bearing on the Issue of Artificial Gravity for Future Manned Space Vehicles. *Aerospace Medicine*, *44*(2), 163–168.

Bles, W., de, G. B., Bos, J. E., Groen, E., & Krol, J. R. (1997). A sustained hyper-g load as a tool to simulate space sickness. *Journal of Gravitational Physiology*, *4*(2), P1-4.

Bles, W., Graaf, B. de, & Bos, J. E. (1996). Space adaptation syndrome (SAS) and sickness induced by centrifugation (SIC): Vestibular consequences of earth anomalous gravity. *Journal of Vestibular Research*, *4 Supplement 1; S*(6), S66.

Bretl, K. N., & Clark, T. K. (2020). Improved feasibility of astronaut short-radius artificial gravity through a 50-day incremental, personalized, vestibular acclimation protocol. *Npj Microgravity*, *6*(1), 22. https://doi.org/10.1038/s41526-020-00112-w

Bretl, K. N., McCusker, A. T., Sherman, S. O., Mitchell, T. R., Dixon, J. B., & Clark, T. K. (2019). Tolerable acclimation to the cross-coupled illusion through a 10-day, incremental, personalized protocol. *Journal of Vestibular Research*, *29*(2–3), 97–110. https://doi.org/10.3233/VES-190656

Buckey, J. C., Lan, M., & Lewis, L. D. (2026). Space travel-associated motion sickness and its treatment. *British Journal of Clinical Pharmacology*, *92*(1), 5–10. https://doi.org/10.1002/bcp.70056

Cheung, C. C., Hecht, H., Jarchow, T., & Young, L. R. (2007). Threshold-based vestibular adaptation to cross-coupled canal stimulation. *Journal of Vestibular Research-Equilibrium & Orientation*, *17*(4), 171–181. (ISI:000257249500003).

Clark, T.K. (2019). Effects of Spaceflight on the Vestibular System. In Pathak, Y., Araujo dos Santos, M., & Zea, L. (Eds.), *Handbook of Space Pharmaceuticals*. Springer.

Clément, G., Moudy, S., Macaulay, T. R., & Wood, S. J. (2025). Effects of short-duration spaceflight on the execution of critical mission tasks. *Frontiers in Physiology*, *16*, 1677377. https://doi.org/10.3389/fphys.2025.1677377

Davis, J. R., Jennings, R. T., & Beck, B. G. (1993). Comparison of treatment strategies for Space Motion Sickness. *Acta Astronautica*, *29*(8), 587–591. https://doi.org/10.1016/0094-5765(93)90074-7

Davis, J. R., Jennings, R. T., Beck, B. G., & Bagian, J. P. (1993). *Treatment Efficacy of Intramuscular Promethazine for Space Motion Sickness*.

Davis, J. R., Vanderploeg, J. M., Santy, P. A., Jennings, R. T., & Stewart, D. F. (1988). *Space Motion Sickness During 24 Flights of the Space Shuttle*.

Diamond, S. G., & Markham, C. H. (1991). Prediciton of Space Motion Sickness Susceptibility by Disconjugate Eye Torsion in Parabolic Flight. *Aviation Space and Environmental Medicine*, *62*(3), 201–205.

Diamond, S. G., & Markham, C. H. (1992). Validating the Hypothesis of Otolith Asymmetry as a Cause of Space Motion Sickness. *Annals of the New York Academy of Sciences*, *656*, 725–731. (ISI:A1992JM22100055).

Diamond, S. G., Markham, C. H., & Money, K. E. (1990). Instability of Ocular Torsion in Zero Gravity—Possible Implications for Space Motion Sickness. *Aviation Space and Environmental Medicine*, *61*(10), 899–905. (ISI:A1990EA36600004).

Golding, J. F. (1998). Motion sickness susceptibility questionnaire revised and its relationship to other forms of sickness. *Brain Research Bulletin*, *47*(5), 507–516. https://doi.org/10.1016/S0361-9230(98)00091-4

Golding, J. F. (2006). Predicting individual differences in motion sickness susceptibility by questionnaire. *Personality and Individual Differences*, *41*(2), 237–248. https://doi.org/10.1016/j.paid.2006.01.012

Golding, J. F. (2016). Motion Sickness. In J. M. Furman & T. Lempert (Eds.), *Handbook of Clinical Neurology* (Vol. 137). Elsevier B.V. https://doi.org/10.1016/B978-0-444-63437-5.00027-3

Gorgiladze, G. I., & Brianov, I. I. (1989). Space motion sickness. *Kosmicheskaia biologiia i aviakosmicheskaia meditsina*, *23*(3), 4–14.

Gray, G., & Johnston, S. L. (2008). Medical Evaluations and Standards. In M. R. Barratt & S. L. Pool (Eds.), *Principles of Clinical Medicine for Space Flight* (1st ed., pp. 59–67). Springer. https://doi.org/10.1007/978-0-387-68164-1_3

Graybiel, A. (1975). Individual differences in susceptibility to motion sickness among six Skylab astronauts. *Acta Astronautica*, *2*, 155–174.

Graybiel, A. (1980). Space Motion Sickness: Skylab Revisited. *Aviation Space and Environmental Medicine*, *51*(8), 814–822.

Graybiel, A., Miller, J. L., & Homick, J. L. (1977). Experiment M131: Human vestibular function. In *Biomedical Results from Skylab* (pp. 74–103). U.S. Government Printing Office.

Graybiel, A., Wood, C. D., Miller, E. F., & Cramer, D. B. (1968). Diagnostic criteria for grading the severity of acute motion sickness. *Aerospace Medicine*, *39*(5), 453–455.

Hardy, J. G., Theriot, C. A., Oswald, T., & Clément, G. (2025). Spaceflight Standard Measures is a multidisciplinary study that systematically monitors risks to astronaut health and performance. *Npj Microgravity*, *11*(1), 78. https://doi.org/10.1038/s41526-025-00532-6

Hecht, H., Brown, E. L., & Young, L. R. (2002). Adapting to artificial gravity (AG) at high rotational speeds. *Journal of Gravitational Physiology: A Journal of the International Society for Gravitational Physiology*, *9*(1), P1-5.

Heer, M., & Paloski, W. H. (2006). Space motion sickness: Incidence, etiology, and countermeasures. *Autonomic Neuroscience*, *129*(1–2), 77–79. https://doi.org/10.1016/j.autneu.2006.07.014

Jarchow, T., & Young, L. R. (2007). Adaptation to head movements during short radius centrifugation. *Acta Astronautica*, *61*(10), 881–888. (ISI:000250431500008). https://doi.org/10.1016/j.actaastro.2006.12.022

Jennings, R. T. (1998). Managing Space Motion Sickness. *Journal of Vestibular Research*, *8*(1), 67–70. https://doi.org/10.3233/VES-1998-8110

Lackner, J. R., & DiZio, P. (2006). Space motion sickness. *Experimental Brain Research*, *175*(3), 377–399. https://doi.org/10.1007/s00221-006-0697-y

Lonner, T. L., Allred, A. R., Bonarrigo, L., Gopinath, A., Smith, K., Kravets, V., Groen, E. L., Oman, C., DiZio, P., Lawson, B. D., & Clark, T. K. (2023). Virtual reality as a countermeasure for astronaut motion sickness during simulated post-flight water landings. *Experimental Brain Research*, *241*(11–12), 2669–2682. https://doi.org/10.1007/s00221-023-06715-5

Lonner, T. L., Allred, A. R., Gopinath, A. R., Morgheim, T., Groen, E. L., Oman, C. M., DiZio, P., Lawson, B. D., Drecksler, S. R., & Clark, T. K. (2025). Reducing motion sickness during simulated astronaut post-spaceflight water landings using anticipatory cues or postural control. *Npj Microgravity*, *11*(1), 21. https://doi.org/10.1038/s41526-025-00478-9

Mark, S., Scott, G. B. I., Donoviel, D. B., Leveton, L. B., Mahoney, E., Charles, J. B., & Siegel, B. (2014). The Impact of Sex and Gender on Adaptation to Space: Executive Summary. *Journal of Women's Health*, *23*(11), 941–947. https://doi.org/10.1089/jwh.2014.4914

Matheson, B. E., Walle, M., Bugbird, A. R., Rosenberg, M., Mateus, J., & Boyd, S. K. (2026). Early skeletal deteriorations following short-duration spaceflight. *Npj Microgravity*, *12*(1), 28. https://doi.org/10.1038/s41526-026-00578-0

Matsnev, E. I., Yakovleva, I. Y., Tarasov, I. K., Alekseev, V. N., Kornilova, L. N., Mateev, A. D., & Gorgiladze, G. I. (1983). Space motion sickness: Phenomenology, countermeasures, and mechanisms. *Journal of Sound and Vibration*, *90*(4), 611. https://doi.org/10.1016/0022-460X(83)90820-9

Menon, A., Mampre, D., Isaacman, J., Poteet, S., Gillis, S., Mateus, J., & Rosenberg, M. (2025). Cardiopulmonary Resuscitation Techniques on the Dragon Spacecraft: A Polaris Dawn Mission Study. *Wilderness & Environmental Medicine*, *36*(1_suppl). https://doi.org/10.1177/10806032251346746

Nicogossian, A., & Uri, J. (1994). Vehicles for human space. In *Space Physiology and Medicine* (3rd ed., pp. 81–108).

Nooij, S. A. E., Bos, J. E., Groen, E. L., Bles, W., & Ockels, W. J. (2007). Space sickness on earth. *Microgravity Science and Technology*, *19*(5), 113–117. https://doi.org/10.1007/BF02919464

Ockels, W. J., Furrer, R., & Messerschmid, E. (1990). Simulation of space adaptation syndrome on earth. *Experimental Brain Research*, *79*(3), 661–663. https://doi.org/10.1007/BF00229334

Oman, C. M. (1987). Spacelab experiments on space motion sickness. *Acta Astronautica*, *15*(1), 55–66. https://doi.org/10.1016/0094-5765(87)90066-X

Oman, C. M. (1990). Motion Sickness—A Synthesis and Evaluation of the Sensory Conflict Theory. *Canadian Journal of Physiology and Pharmacology*, *68*(2), 294–303. (ISI:A1990CR05700023).

Oman, C. M. (1998). Sensory Conflict Theory and Space Sickness: Our Changing Perspective. *Journal of Vestibular Research*, *8*(1), 51–56.

Oman, C. M. (2002). Neurovestibular Adaptation to Spaceflight: Research Progress. *Journal of Vestibular Research*, *12*(5–6), 201–203. (ISI:000185978800001).

Oman, C. M., Lichtenberg, B. K., Money, K. E., & McCoy, R. K. (1986). M.I.T./Canadian vestibular experiments on the Spacelab-1 mission: 4. Space motion sickness: symptoms, stimuli, and predictability. *Experimental Brain Research*, *64*(2). https://doi.org/10.1007/BF00237749

Ortega, H. J., & Harm, D. L. (2019). Space and entry motion sickness. In *Principles of Clinical Medicine for Space Flight* (pp. 441–456). (Original work published Springer)

Ortega, H. J., Harm, D. L., & Reschke, M. F. (2019). Space and Entry Motion Sickness. In M. R. Barratt, E. S. Baker, & S. L. Pool (Eds.), *Principles of Clinical Medicine for Space Flight* (pp. 441–456). Springer New York. https://doi.org/10.1007/978-1-4939-9889-0_14

Putcha, L., Younker, D., & Daniels, V. (2012). Space Motion Sickness—Analysis of Medical Debriefs Data for Incidence and Treatment. *83rd AsMA Annual Scientific Meeting*. https://ntrs.nasa.gov/citations/20110020471

Ramsburg, C. F., Wood, S. J., Lackner, J. R., Moynihan, S., Reschke, M. F., Denise, P., & Clément, G. (2025). Potential benefits and human systems integration of parastronauts with bilateral vestibulopathy for a space mission. *Frontiers in Neurology*, *16*, 1556553. https://doi.org/10.3389/fneur.2025.1556553

Reschke, M. F., Bloomberg, J. J., Harm, D. L., Paloski, W. H., Layne, C., & McDonald, V. (1998). Posture, locomotion, spatial orientation, and motion sickness as a function of space flight. *Brain Research Reviews*, *28*(1–2), 102–117. (ISI:000077147400014). https://doi.org/10.1016/S0165-0173(98)00031-9

Reschke, M. F., Cohen, H. S., Cerisano, J. M., Clayton, J. A., Cromwell, R., Danielson, R. W., Hwang, E. Y., Tingen, C., Allen, J. R., & Tomko, D. L. (2014). Effects of Sex and Gender on Adaptation to Space: Neurosensory Systems. *Journal of Womens Health*, *23*(11), 959–962. (ISI:000344868200013). https://doi.org/10.1089/jwh.2014.4908

Reschke, M. F., Good, E. F., & Clément, G. R. (2017). Neurovestibular Symptoms in Astronauts Immediately after *Space Shuttle* and *International Space Station* Missions. *OTO Open*, *1*(4), 2473974X17738767. https://doi.org/10.1177/2473974X17738767

Reschke, M. F., Wood, S. J., & Clément, G. R. (2018). A Case Study of Severe Space Motion Sickness. *Aerospace Medicine and Human Performance*, *89*(8), 749–753. https://doi.org/10.3357/AMHP.5071.2018

Thornton, W. E., & Bonato, F. (2013). Space Motion Sickness and Motion Sickness: Symptoms and Etiology. *Aviation, Space, and Environmental Medicine*, *84*(7), 716–721. https://doi.org/10.3357/ASEM.3449.2013

Thornton, W. E., Moore, T. P., Pool, S. L., & Vanderploeg, J. M. (1987). Clinical Characterization and Etiology of, Space Motion Sickness. *Aviation Space and Environmental Medicine*, *58*(9, Suppl.:A), 1–8.

Weerts, A. P., Vanspauwen, R., Fransen, E., Jorens, P. G., Van De Heyning, P. H., & Wuyts, F. L. (2014). Space Motion Sickness Countermeasures: A Pharmacological Double-Blind, Placebo-Controlled Study. *Aviation, Space, and Environmental Medicine*, *85*(6), 638–644. https://doi.org/10.3357/ASEM.3865.2014

Yates, B. J., Miller, A. D., & Lucot, J. B. (1998). Physiological basis and pharmacology of motion sickness: An update. *Brain Res Bull*, *47*(5), 395–406.

Young, L. R., Hecht, H., Lyne, L. E., Sienko, K. H., Cheung, C. C., & Kavelaars, J. (2001). Artificial Gravity: Head Movements during Short-Radius Centrifugation. *Acta Astronautica*, *49*(3–10), 215–226. (ISI:000170751900014).

Young, L. R., Oman, C. M., Watt, D. G. D., Money, K. E., & Lichtenberg, B. K. (1984a). Spatial Orientation in Weightlessness and Readaptation to Earth's Gravity. *Science*, *225*(4658), 205–208. (ISI:A1984SY85400049). https://doi.org/10.1126/science.6610215

Young, L. R., Oman, C. M., Watt, D. G. D., Money, K. E., & Lichtenberg, B. K. (1984b). Spatial Orientation in Weightlessness and Readaptation to Earth's Gravity. *Science*, *225*(4658), 205–208.

Young, L. R., Oman, C. M., Watt, D. G. D., Money, K. E., Lichtenberg, B. K., Kenyon, R. V., & Arrott, A. P. (1986). Mit Canadian Vestibular Experiments on the Spacelab-1 Mission .1. Sensory Adaptation to Weightlessness and Readaptation to One-G - an Overview. *Experimental Brain Research*, *64*(2), 291–298. (ISI:A1986E648100008).